\documentclass[onecolumn,linenumbers]{aastex631}
\usepackage{multirow}
\usepackage{enumerate}
\usepackage{diagbox}
\usepackage{enumitem}
\usepackage{amssymb}
\usepackage{amsmath}
\usepackage{lineno}
\usepackage{bm,bbm}

\nolinenumbers

\shorttitle{Time delay of FRB population with respect to SFH}
\shortauthors{H.-N. Lin, X.-Y. Li and R. Zou}
\graphicspath{{./}{figures/}}

\begin{document}

\title{Time delay of fast radio burst population with respect to the star formation history}

\correspondingauthor{Hai-Nan Lin}\email{linhn@cqu.edu.cn}

\author[0000-0003-1659-3368]{Hai-Nan Lin}
\affiliation{Department of Physics, Chongqing University, Chongqing 401331, China}
\affiliation{Chongqing Key Laboratory for Strongly Coupled Physics, Chongqing University, Chongqing 401331, China}

\author[0009-0005-2843-2360]{Xin-Yi Li}
\affiliation{Department of Physics, Chongqing University, Chongqing 401331, China}
\affiliation{Chongqing Key Laboratory for Strongly Coupled Physics, Chongqing University, Chongqing 401331, China}

\author[0000-0002-9998-6222]{Rui Zou}
\affiliation{Department of Physics, Chongqing University, Chongqing 401331, China}
\affiliation{Chongqing Key Laboratory for Strongly Coupled Physics, Chongqing University, Chongqing 401331, China}

\begin{abstract}
\nolinenumbers
In spite of significant progress in the research of fast radio bursts (FRBs) in recent decade, their origin is still under extensive debate. Investigation on the population of FRBs can provide new insight into this interesting problem. In this paper, based on the first CHIME/FRB catalog, we construct a Bayesian framework to analyze the FRB population, with the selection effect of the CHIME telescope being properly taken into account. The energy function is modeled as the power-law with an exponential cutoff. Four redshift distribution models are considered, i.e., the star formation history (SFH) model, and three time-delayed models (Gaussian delay, log-normal delay, and power-law delay). The free parameters are simultaneously constrained using Bayesian inference method, and the Bayesian information criterion (BIC) is used in model comparison. According to BIC, the log-normal delay model fits the data best. The power-law delay model and Gaussian delay model can also give reasonable fits, although they are not as good as the log-normal delay model. However, the SFH model is strongly disfavored compared with the three time-delayed models. The energy function is tightly constrained and is almost independent of the redshift models, with the best-fitting power-law index $\alpha\approx 1.8$, and cut-off energy $\log(E_c/{\rm erg})\approx 42$. The FRB population shows on average $3\sim 5$ billion years time delay with respect to the SFH. Therefore, the hypothesis that the FRB population traces the SFH is conclusively ruled out.
\end{abstract}

\section{Introduction}\label{sec:introduction}

Fast radio bursts (FRBs) are energetic (up to $\sim 10^{43}$ erg) and short-duration ($\sim$ milliseconds) radio pulses coming from the Universe, see e.g. \citep{Platts:2018hiy,Petroff:2019tty,Xiao:2021omr,Zhang_2023} for recent review. The first FRB (which is now named FRB 20010724) was found from the archival data of the 64-metre Parkes radio telescope in 2007 \citep{Lorimer:2007qn}, but it had not aroused people's attention until four other similar signals was discovered six years later \citep{Thornton:2013iua}. Thereafter, researches in FRB fields have achieved rapid progress, especially in recent decade. The localization of host galaxy and the direct measurement of redshift confirm the cosmological origin \citep{Keane:2016yyk,Chatterjee:2017dqg}. Phenomenologically, FRBs can be divided into two categories, i.e., repeater and non-repeater. In 2016, the  Arecibo 305-metre radio telescope detected several similar radio signals from the same sky position of FRB 20121102 \citep{Spitler:2016dmz}, making it to be the first discovered repeating FRB. Later on, it was found that the source of FRB 20121102 is extremely active, which emits more than one thousand bursts in about 60 hours spanning less than two months \citep{Li:2021hpl}. With the operation of the Canadian Hydrogen Intensity Mapping Experiment (CHIME, \citealt{CHIMEFRB:2018mlh}), we are able to detect FRBs in real time, thus extensively enlarging the FRB sample \citep{CHIMEFRB:2021srp}. Up to now, about eight hundred FRB sources have been published \citep{Xu:2023did}, a majority of which are discovered by the CHIME telescope. With the join of high-sensitive radio telescopes such as the Square Kilometre Array (SKA) and the Five-hundred-meter Aperture Spherical radio Telescope (FAST), we expect that the number of FRB sources, as well as the number of bursts of individual repeating sources will significantly increase in the near future.

Although hundreds of FRBs have been found, most of them are not well localized, leaving their origins largely unknown. But there is no dispute that most FRBs are extragalactic origin, because the observed dispersion measure (DM) greatly excesses the maximum contribution from the Milky Way. Only one event (FRB 20200428)\footnote{This burst is now renamed as FRB 20200428D, since some other FRBs are observed on the same day.} is confirmed to originate from the Milky Way. This specific burst is found to be associated with a short X-ray burst from the Galactic magnetar SGR 1935+2154 \citep{Andersen:2020hvz,Bochenek:2020zxn}. This implies that at least some, if not all FRBs originate from magnetars. Although young magnetars are proposed as the leading source candidates, recent observations suggest that there may be multiple FRB progenitors \citep{moroianu2022assessment}. Similar to the classification of FRBs themselves, FRB progenitors can also be classified into non-repeater and repeater categories, thus leading to two classes of FRB models: catastrophic models for the non-repeaters and non-catastrophic models for the repeaters. Initially, all FRBs were considered as non-repeaters and are most likely catastrophic events, until the discovery of the first repeating event FRB 20121102 \citep{Spitler:2016dmz}. Given the significantly higher event rate density of FRBs than the catastrophic events in the Universe \citep{Ravi:2019alc,Luo:2020wfx}, it is highly probable that a majority of FRBs originate from repeating sources. So far, at least dozens of FRB models have been proposed, see e.g. \citet{Platts:2018hiy} for excellent review. Considering the short-duration and extremely energetic properties of FRBs, the most popular FRB models involve at least one compact object, such as neutron star, magnetar or white dwarf in the center of FRB source. For instance, the merger of binary white dwarfs \citep{Kashiyama2013}, the merger of binary neutron stars \citep{Wang:2016dgs,Totani:2013lia}, the coalescence of neutron star -- black hole systems \citep{Mingarelli:2015bpo}, the collapse of neutron star to black hole or quark star \citep{Falcke:2013xpa,Shand:2015uda}, the starquake of pulsar \citep{Wang_2018}, just name a few.

The physical origin of FRBs is still under extensive debate, and there is no conclusive evidence for preferring one model against the others. Statistical analysis shows that repeaters and non-repeaters have very different properties, hence may have different progenitors \citep{Pleunis_2021}. Different progenitors will lead to different FRB population distributions. Therefore, investigating the population of FRBs is of great importance in order to distinguish different FRB models. Figure 1 of \citet{Zhang:2020ass} gives a comprehensive summary of the most popular FRB models and the corresponding population distributions. For instance, the core-collapse models predict that the FRB population should trace the star formation history (SFH), the models involving the merger of binary compact objects predict that FRB population should have a significant time delay with respect to the SFH, while for some other models (such as the cosmic comb model and cosmic string model) the FRB population is unpredictable.

The population of FRBs have been extensively studied in literature. Based on two FRB samples detected by the Parkes and the Australian Square Kilometre Array Pathfinder (ASKAP), \citet{Zhang:2020ass} found that the SFH model and three time-delayed models can all match the data well, as long as the free parameters are chosen properly. Using similar samples, \citet{James:2021oep} showed that FRB population evolves with redshift in a way consistent with, or faster than the SFH. The large FRB sample observed by the CHIME telescope \citep{CHIMEFRB:2021srp} allows us to study the FRB population in details. Based on the first CHIME/FRB catalog, \citet{zhang2022chime} showed that the data can well match the time-delayed models, but the SFH model is conclusively excluded. It is also possible that the CHIME/FRB sample is a hybrid sample consists of both the time-delayed population and SFH population. Based on the same CHIME/FRB sample, \cite{Qiang:2021ljr} investigated the SFH model as well as some empirical models, and also arrived at the conclusion that the SFH model is ruled out. Very recently, \citet{lin2023revised} revised the constraints on the FRB population using the first CHIME/FRB catalog, and found that more than one models can match the data if the parameters are optimized, but the SFH model is strongly disfavored. All these studies rule out the hypothesis that FRBs trace the SFH, but the concrete FRB population remains to be unconstrained.

The observed FRB population is mainly influenced by two intrinsic distributions of FRB sources, i.e. the energy distribution and redshift distribution. Several independent works show that the intrinsic energy distribution follows the power-law, and maybe with a high-energy exponential cutoff, which is independent of the redshift distribution \citep{Luo:2018tiy,Luo:2020wfx,lu2020unified,lin2023revised}. Compared with the energy distribution, the redshift distribution is less known. Different progenitor theories predict different redshift distributions. The most motivated redshift distribution model is the SFH model, which predicts that the occurrence rate of FRBs should mirror the cosmic SFH. However, the SFH model has already been ruled out by observations \citep{zhang2022chime,Qiang:2021ljr,lin2023revised}. Due to the additional time delay caused by gravitational wave radiation in the mergers of binary compact objects, this particular channel predicts a redshift distribution that deviates significantly from that of the SFH model. Therefore, the time-delayed models have been proposed \citep{Zhang:2020ass,zhang2022chime}. Using the Kolmogorov-Smirnov (KS) test method, it has been shown that the time-delayed models can to some extent match a large sample of FRBs, as long as the model parameters are chosen properly \citep{zhang2022chime}. However, the KS-test method used in \citet{zhang2022chime} relies on burdensome simulations, and the optimal parameters couldn't be acquired.

The recently published CHIME/FRB catalog \citep{CHIMEFRB:2021srp} offers a large FRB sample, facilitating an in-depth study of the FRB population. In this paper, we reinvestigate the FRB population use the first CHIME/FRB catalog. Three time-delayed redshift distribution models are considered: Gaussian delay, log-normal delay and power-law delay. These models have already been used to study the population of FRBs \citep{Zhang:2020ass,zhang2022chime} and gamma-ray bursts \citep{Wanderman_2015,Sun_2015}. The SFH model is also considered for comparison, although is has already been ruled out. With the selection effect of telescope taking into account, we construct a joint likelihood of energy, fluence and redshift. The model parameters are optimized using the Bayesian inference method, and the Bayesian information criterion (BIC) are used in model comparison.

The structure of the rest parts of this paper are organized as follows: In Section \ref{sec:method}, we introduce the FRB population models and construct a Bayesian framework in order to constrained the models parameters. In Section \ref{sec:results}, based on the Bayesian framework, the model parameters are constrained using the data of the first CHIME/FRBs catalog. Finally, relevant discussion and conclusions are given in Section \ref{sec:conclusions}.

\section{Methodology}\label{sec:method}

The redshift of FRB is necessary to investigate the population. Unfortunately, a majority of FRBs have no direct measurement of redshift. One way to solve this problem is using the dispersion measure (DM) as a proxy of redshift. DM is the integral of the number density of free electrons along line-of-sight, which can be extracted directly from the dynamical spectra. FRBs at further distance are expected to confront more electrons thus have larger DMs. The observed DM of an extragalactic FRB, can in general be decomposed into four terms \citep{Deng:2013aga,Gao:2014iva,Macquart:2020lln},
\begin{equation}\label{eq:DM_obs}
    {\rm DM_{obs}}={\rm DM_{MW,ISM}}+{\rm DM_{WM,halo}}+{\rm DM_{IGM}}+\frac{{\rm DM_{host}}}{1+z}.
\end{equation}
The first term, $\mathrm{DM_{MW,ISM}}$, is the contribution from the Milky Way interstellar medium, which can be estimated using Milky Way electron density models, such as the NE2001 model \citep{Cordes:2002wz} and YMW16 model \citep{Yao_2017msh}. The second term, $\mathrm{DM_{WM,halo}}$, represents the contribution from the Milky Way halo. This term has not been well constrained yet, but it is approximately estimated to be $\sim 50~{\rm pc~cm}^{-3}$ \citep{Prochaska:2019mkd,Macquart:2020lln}. The third term, ${\rm DM_{IGM}}$, is the contribution from the intergalactic medium (IGM), which can be calculated given a specific cosmological model. Finally, the last term $\rm DM_{host}$ represents the dispersion caused by free electrons in the FRB host galaxy.

In the standard ${\rm\Lambda}$CDM model, the average value of the ${\rm DM_{IGM}}$ term can be calculated as \citep{Deng:2013aga}
\begin{equation}\label{eq:DM_IGM}
    \langle{\rm DM_{IGM}}(z)\rangle=\frac{3cH_0\Omega_bf_{\rm IGM}f_{e}}{8\pi Gm_p}\int_0^z\frac{1+z}{\sqrt{\Omega_m(1+z)^3+\Omega_\Lambda}}dz,
\end{equation}
where $c$ is the speed of light, $m_p$ is the proton mass, $G$ is the Newtonian gravitational constant, $f_{\rm IGM}=0.84$ is the baryon mass fraction in IGM, and $ f_e=7/8$ is the electron fraction. The Hubble constant $H_0$, the vacuum energy density $\Omega_\lambda$ and the baryon density $\Omega_b$ are fixed to the Planck 2018 results \citep{Aghanim:2018eyx}. Note that the actual value of ${\rm DM_{IGM}}$ may significantly deviate from equation (\ref{eq:DM_IGM}) due to the fluctuation of matter density. Numerical simulations demonstrate that the distribution of $\rm DM_{IGM}$ can be well modeled by the quasi-Gaussian function \citep{Macquart:2020lln,Zhang:2020xoc},
\begin{equation}
	p_{\rm IGM}(\Delta)=A\Delta^{-\beta}\exp\left[-\frac{(\Delta^{-\alpha}-C_0)^2}{2\alpha^2\sigma_{\rm IGM}^2}\right], ~~~\Delta>0,
\end{equation}
where $\rm \Delta \equiv DM_{IGM}/\left \langle DM_{IGM} \right \rangle$, $\alpha = \beta = 3$, the effective standard deviation $\sigma_{\rm IGM}=Fz^{-1/2}$ with $F$ being a free parameter, $A$ is the normalization constant, and $C_0$ is chosen to ensure that the mean value of $\Delta$ is unity.

Currently, the host term $\rm DM_{host}$ is still poorly known, primarily due to the lack of comprehensive observations of the local surroundings of most FRB sources. The value of $\rm DM_{host}$ can vary significantly from burst to burst, ranging from several tens to several hundreds pc cm$^{-3}$ \citep{Niu_2022,Kirsten:2021llv}. To account for such a large variation, it is usually modeled using the log-normal distribution \citep{Macquart:2020lln,Zhang:2020mgq},
\begin{equation}\label{eq:P_host}
    p_{\rm host}({\rm DM_{host}}|\mu,\sigma_{\rm host})=\frac{1}{\sqrt{2\pi}{\rm DM_{host}}\sigma_{\rm host}} \exp\left[-\frac{(\ln {\rm DM_{host}}-\mu)^2}{2\sigma_{\rm host}^2}\right],
\end{equation}
where the two free parameters $\mu$ and $\sigma_{\rm host}$ are the mean and standard deviation of ln $\rm DM_{host}$, respectively.

The first two terms of the right-hand-side of equation (\ref{eq:DM_obs}) can be subtracted from the total observed $\mathrm{DM_{obs}}$, leaving behind the extragalactic DM,
\begin{equation}\label{eq:DM_E}
	{\rm DM_E}\equiv {\rm DM_{obs}}-{\rm DM_{MW,ISM}}-{\rm DM_{WM,halo}}={\rm DM_{IGM}}+\frac{{\rm DM_{host}}}{1+z}.
\end{equation}
According to the probability theory, the distribution of ${\rm DM_E}$ can be written as the product of the distributions of $\rm DM_{IGM}$ and $\rm DM_{host}$, and then marginalize over either term \citep{Macquart:2020lln},
\begin{equation}\label{eq:P_E}
    p_E({\rm DM_E}|z)=\int_0^{(1+z)\rm DM_E}p_{\rm host}({\rm DM_{host}}|\mu,\sigma_{\rm host})p_{\rm IGM}({\rm DM_E}-\frac{\rm DM_{host}}{1+z}|F,z)d{\rm DM_{host}},
\end{equation}
where the free parameters $F$, $\mu$ and $\sigma_{\rm host}$ can be tightly constrained using 17 well-localized FRBs \citep{Tang_2023}.

Based on equation (\ref{eq:P_E}), we can reconstruct the ${\rm DM_E}-z$ relation, as is shown in Figure 2 of \citet{Tang_2023}. Then the ${\rm DM_E}-z$ relation can be used to infer the redshift and calculate the energy of FRBs. Given the observed fluence $(F_{\nu})$, the isotropic energy can be calculated as \citep{Macquart:2018jlq,James:2021jbo}
\begin{equation}\label{eq:E}
	E=\frac{4\pi d_L^2}{(1+z)^{2+\beta}}F_{\nu} \Delta\nu,
\end{equation}
where $\beta$ is the spectrum index ($F_\nu\propto \nu^\beta$), $d_L$ is the luminosity distance, and $\Delta\nu$ is the bandwidth in which the FRB is detected. The energies of the first CHIME/FRB catalog are listed in the appendix of \citet{Tang_2023}.

Now we have the distributions of fluence, energy and redshift. These three observables are used simultaneously to constrain the FRB population. The distribution of energy can be well modeled by power-law with a high-energy exponential cutoff \citep{Luo:2018tiy,lu2020unified},
\begin{equation}\label{eq:CPL}
	p(E)\propto \left({\frac{E}{E_c}} \right)^{-\alpha} \exp\left(-\frac{E}{E_c}\right).
\end{equation}
Although the cutoff energy $E_c$ is not well constrained yet, the power-low index $\alpha$ is typically confined to be in the range $1.8 \lesssim \alpha \lesssim 2.0 $ \citep{Luo:2018tiy,Lu_2019,lu2020unified}. Here we treat both $E_c$ and $\alpha$ as free parameters.

The redshift distribution of FRBs in the observer frame is related to the intrinsic event rate density $dN/dtdV$ by \citep{Zhang:2020ass,zhang2022chime}
\begin{equation}\label{eq:redshift}
	p(z)\propto \frac{dt}{dt_{\rm obs}}\frac{dN}{dtdV}\frac{dV}{dz},
\end{equation}
where $dt/dt_{\rm obs} = (1+z)^{-1}$ accounts for the time dilation, and in the flat $\rm \Lambda$CDM model the redshift-dependent specific comoving volume $dV/dz$ is given by
\begin{equation}	
	\frac{dV}{dz}= \frac{4\pi cd^{2}_{C}}{H_0\sqrt{\Omega_m (1+z)^3 + \Omega_{\Lambda}}},
\end{equation}
where $d_C=d_L/(1+z)$ is the comoving distance.

The intrinsic event rate density $dN/dtdV$ of FRB is poorly constrained. The most well-motivated model is the SFH of the Universe. The SFH has been extensively studied, and many parametric forms have been proposed \citep{Yuksel:2008cu,Madau_2014,Madau:2016jbv}. Here we use the updated version of two-segment empirical SFH model \citep{Madau:2016jbv},
\begin{equation}\label{eq:SFH}
	{\rm SFH}=\frac{(1+z)^{2.6}}{1+((1+z)/3.2)^{6.2}}.
\end{equation}
If FRBs originate from core-collapse of massive stars, then the FRB population is expected to follow the SFH. On the other hand, if FRBs originate from the merger of compact binaries, the time delay with respect to SFH is expected, because a binary system must go through an extended inspiral phase prior to eventual coalescence. Assuming that the time delay follows the probability distribution $f(\tau)$, then the intrinsic event rate density of FRBs is the convolution of the SFH with $f(\tau)$ \citep{Wanderman_2015},
\begin{equation}
	\frac{dN}{dtdV} \propto \int_{z}^{\infty} {\rm SFH}(z') \cdot f(t(z) - t(z')) \frac{dt}{dz'}dz' ,
\end{equation}
where the cosmic time is related to redshift through
\begin{equation}
	t(z) = \int_{z}^{\infty} \frac{1}{H_0(1+z)\sqrt{\Omega_m (1+z)^3 + \Omega_{\Lambda}}}dz.
\end{equation}

The distribution of time delay $f(\tau)$ is still unclear. Here, we follow \citet{Zhang:2020ass} and consider three different distributions:

$\bullet$ Gaussian delay model:
\begin{equation}\label{eq_Gd}
	    f(\tau) = \frac{1}{{\sigma \sqrt{2\pi}}} \exp\left(-\frac{\left(\tau-\tau_0\right)^2}{2\sigma^2}\right).
\end{equation}

$\bullet$ Log-normal delay model:
\begin{equation}\label{eq_Lognd}
	    f(\tau) = \frac{1}{{\tau \sigma \sqrt{2\pi}}} \exp\left(-\frac{\left(\ln\tau-\ln\tau_0\right)^2}{2\sigma^2}\right).
\end{equation}

$\bullet$ Power-law delay model:
\begin{equation}\label{eq_Pld}
        f(\tau) = \left(\frac{1-\alpha_\tau}{{\tau_{\rm max}^{1-\alpha_{\tau}}-\tau_{\rm min}^{1-\alpha_{\tau}}}}\right) {\tau}^{-\alpha_\tau}.
\end{equation}
In the power-law delay model, $\tau_{\rm max}$ is the maximum merger delay timescale and $\tau_{\rm min}$ is the minimum. We consider $\tau_{\rm max}=1/H_0$ to be the Hubble time and set $\tau_{\rm min}=20$ Myr. The above three time-delay models have been successfully applied in modeling the redshift distribution of the non-collapsar short gamma-ray bursts \citep{Wanderman_2015}.

Equations (\ref{eq:CPL}) and (\ref{eq:redshift}) are valid if all FRBs are detectable. In the actual case, however, the distributions of energy and redshift observed by a specific telescope strongly depend on the selection effect of the telescope. The selection effect of the CHIME telescope is hard to determine. Here we follow \citet{zhang2022chime} and model the detection efficiency as a function of the specific fluence,
\begin{equation}
    \eta_{\rm det}(F_\nu)=\left( \frac{\mathrm{log}F_\nu-\mathrm{log}F^{\rm min}_{\nu,\rm th}}{\mathrm{log}F_{\nu,\rm th}^{\rm max}-\mathrm{log}F^{\rm min}_{\nu,\rm th}} \right)^n,\quad \mathrm{log}F^{\rm min}_{\nu,\rm th}<\mathrm{log}F^{}_{\nu}<\mathrm{log}F_{\nu,\rm th}^{\rm max}.
\end{equation}
This model assumes a `grey zone' between the minimum specific threshold fluence $F_{\nu,\rm th}^{\rm min}$, and the maximum specific threshold fluence $F_{\nu,\rm th}^{\rm max}$. Within this range FRBs may not be fully detected. FRBs with $F_\nu<F_{\nu,\rm th}^{\rm min}$ are undetectable hence $\eta_{\rm det}=0$, while above $F_{\nu,\rm th}^{\rm max}$ all FRBs are detectable hence $\eta_{\rm det}=1$. Following \citet{zhang2022chime}, we fix $\mathrm{log}F^{\rm min}_{\nu,\rm th}=-0.5$ (as shown by the data), while $\mathrm{log}{F_{\nu,\rm th}^{\rm max}}$ is treated as a free parameter.

By inverting equation (\ref{eq:E}), the specific fluence $F_{\nu}$ can be expressed as a function of energy and redshift,
\begin{equation}\label{eq:F}
	F_\nu(E,z)=\frac{(1+z)^{2+\beta}}{4\pi d_L^2\Delta\nu}E.
\end{equation}
Note that the luminosity distance $d_L(z)$ is a function of redshift. Taking the selection effect into account, the distributions of fluence, energy and redshift observed by the CHIME telescope can be written as \citep{lin2023revised}
\begin{equation}\label{eq_1}
	p_{\rm det}(F_\nu)\propto\frac{d}{dF_\nu}\iint\limits_{F_\nu(E,z)\le F_\nu}p(z)p(E)\eta_{\rm det} (F_\nu(E,z))dE\,dz,
\end{equation}
\begin{equation}\label{eq_2}
	p_{\rm det}(E)\propto\int_{z_{\rm min}}^{z_{\rm max}}p(z)p(E)\eta_{\rm det} (F_\nu(E,z))dz,
\end{equation}
\begin{equation}\label{eq_3}
	p_{\rm det}(z)\propto\int_{E_{\rm min}}^{E_{\rm max}}p(z)p(E)\eta_{\rm det} (F_\nu(E,z))dE.
\end{equation}

Equations (\ref{eq_1},\ref{eq_2},\ref{eq_3}) are the probability density functions (PDFs) of fluence, energy and redshift, respectively. They can be used to fit to the data of the CHIME/FRBs catalog. But in order to avoid the arbitrariness of binning, we use the cumulative distribution functions (CDFs) instead of the PDFs in the fitting procedure. The CDF of a quantity $Q$ is defined by
\begin{equation}\label{eq:cdf}
    N_\mathrm{det}(>Q) = A\int_{Q}^{Q_\mathrm{max}}p_\mathrm{det}(Q) dQ, ~~~ Q = \left\{F_\nu,\, E,\, z\right\},
\end{equation}
where the normalization factor is chosen to match the total number of data points. The $\chi^2$ of quantity $Q$ can be written as
\begin{equation}
  \chi^2_Q = \sum_{i}^{n} \frac{[N_\mathrm{det}(>Q_i) - N(>Q_i)]^2}{\sigma_{i}^2}, ~~~ Q = \left\{F_\nu,\, E,\, z\right\} ,
\end{equation}
where $N(>Q_i)$ is the number of FRBs whose $F_\nu$, $E$ or $z$ is larger than that of the $i$-th FRB, and the uncertainty is given by $\sigma_i = \sqrt{N(>Q_i)}$ \citep{Aschwanden:2015}.
The total $\chi^2$ is constructed as
\begin{equation}\label{chi2total}
  \chi^2_{\mathrm{total}} = \chi^2_{F_\nu} + \chi^2_z + \chi^2_E.
\end{equation}
So the likelihood of observing a sample of FRBs with $F_\nu$, $E$ and $z$ can be written as
\begin{equation}\label{eq:likelihood}
  \mathcal{L}(\mathrm{FRBs|{\bm\theta}}) \propto \mathrm{exp}\left(-\frac{1}{2}\chi^2_{\mathrm{total}}\right),
\end{equation}
where ${\bm\theta}$ is the set of free parameters, including $\alpha$ and $E_c$ in the energy distribution, $\log F_{\nu,\rm th}^{\rm max}$ and $n$ in the selection function, and the parameters in the time-delayed models as specified in equations (\ref{eq_Gd},\ref{eq_Lognd},\ref{eq_Pld}). Then the posterior PDFs of free parameters can be obtained according to the Bayesian theorem,
\begin{equation}\label{posterior}
  P(\mathrm{{\bm\theta}|FRBs})\propto\mathcal{L}(\mathrm{FRBs|{\bm\theta}})P_0({\bm\theta}),
\end{equation}
where $P_0$ is the prior of the parameters.

\section{Data and results}\label{sec:results}

In this section, we use the first CHIME/FRB catalog \citep{CHIMEFRB:2021srp} to constrain the free parameters. The first CHIME/FRB catalog  contains 536 bursts in total, including 474 apparently non-repeating bursts and 62 repeating bursts from 18 different FRB sources. All of the bursts have well measured $\rm DM_{obs}$, but most of them have no redshift measurement. We employ the ${\rm DM_E} - z$ relation to infer the redshift and calculate the energy. In contrast to some previous works (such as \citealt{zhang2022chime}, \citealt{Qiang:2021ljr} and \citealt{Shin:2022crt}) which combined the non-repeating bursts and the first burst from the repeating FRB sources, here we only select the non-repeating bursts. There are many reasons why we do this so. First, repeating and non-repeating FRBs may have different origins, so their population may be different. Including the repeating FRBs into our sample may mingle the FRB population. Second, for the repeating FRBs, burst energies from the same FRB source vary significantly, so there is no unique way to define the FRB energy. Third, in the first CHIME/FRB catalog, there are only 18 repeating FRB sources, in contrast to more than four hundred apparently non-repeating FRB sources. We expect that including the repeating FRBs into our data sample will not significantly affect the results. But be caution that (even all) apparently non-repeaters could become repeaters in the future. In addition, we only consider the FRBs with $\rm DM_E > 100$ pc cm$^{-3}$ (in total 444 FRBs). This is because the ${\rm DM_{host}}$ term is expected to be in the order of magnitude $100~{\rm pc~cm^{-3}}$. For FRBs with $\rm DM_E < 100$ pc cm$^{-3}$, the contribution of the $\rm DM_{host}$ term dominates over the $\rm DM_{IGM}$ term, so the inferred redshift may be strongly biased. Three FRBs with inferred redshift $z_{\rm inf}>3$ are excluded, because the ${\rm DM_E}-z$ relation is obtained based on the assumption that hydrogen and helium are fully ionized, which may be broken at $z>3$. Excluding five FRBs with no measurement of fluence, the remaining 436 bursts are call the ``Full sample" for convenience. The inferred redshifts and energies of the Full sample are listed in \citet{Tang_2023}.

\begin{figure}[htbp]
	\centering
	\includegraphics[width=0.32\textwidth]{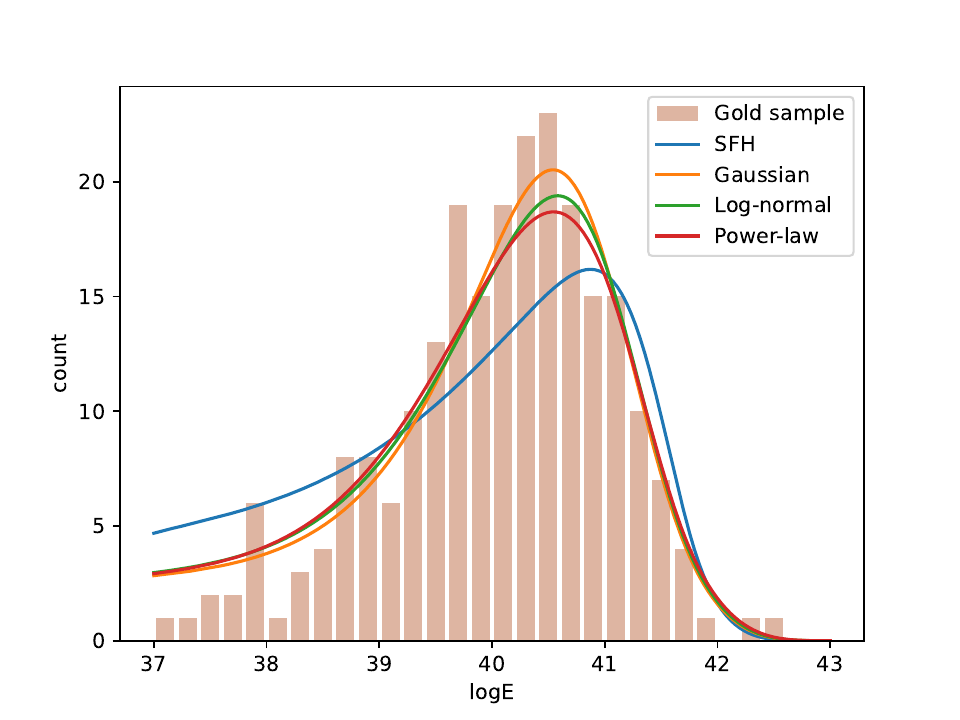}
	\includegraphics[width=0.32\textwidth]{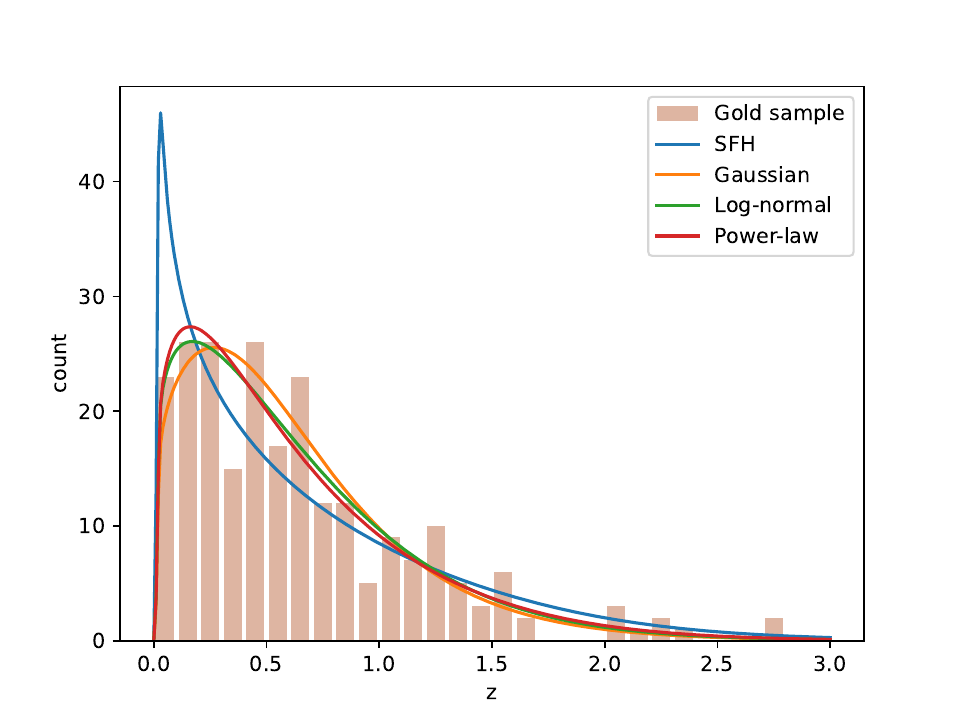}
	\includegraphics[width=0.32\textwidth]{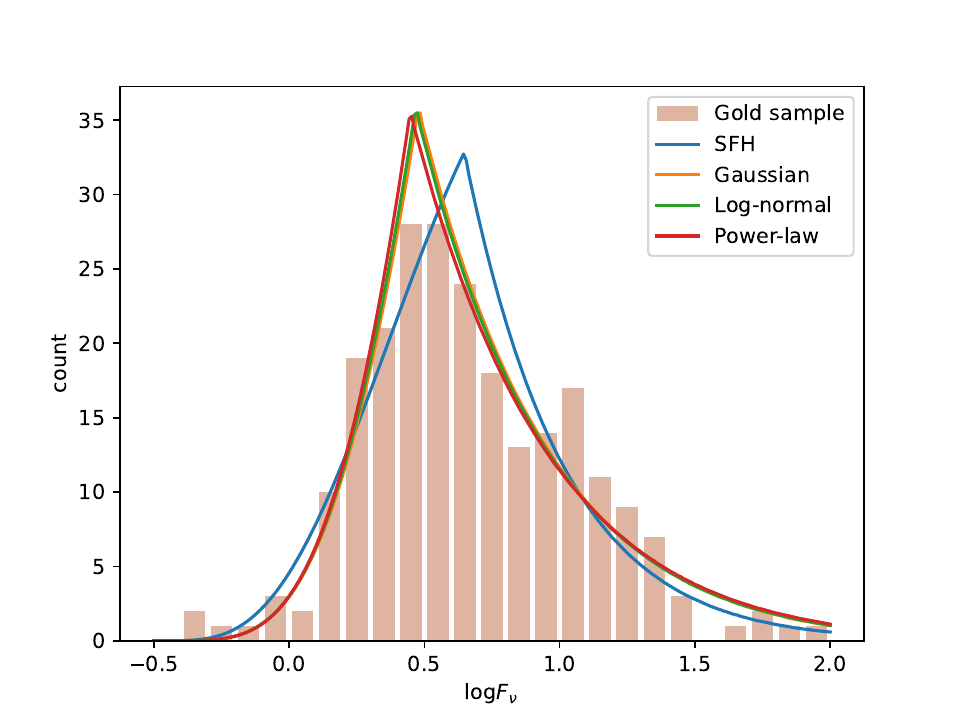}
    \includegraphics[width=0.32\textwidth]{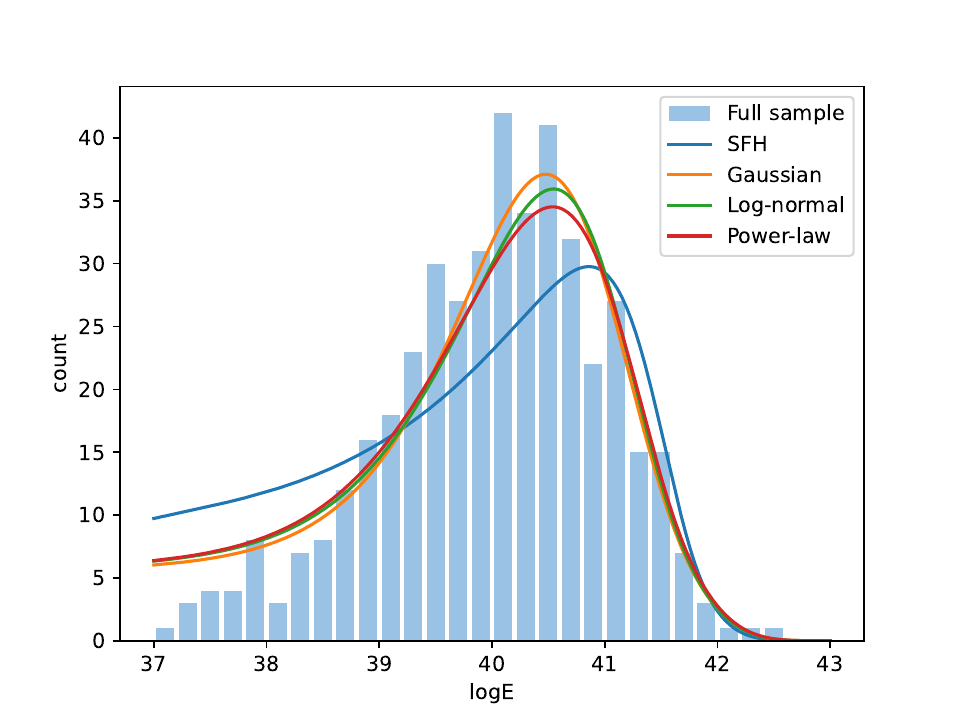}
	\includegraphics[width=0.32\textwidth]{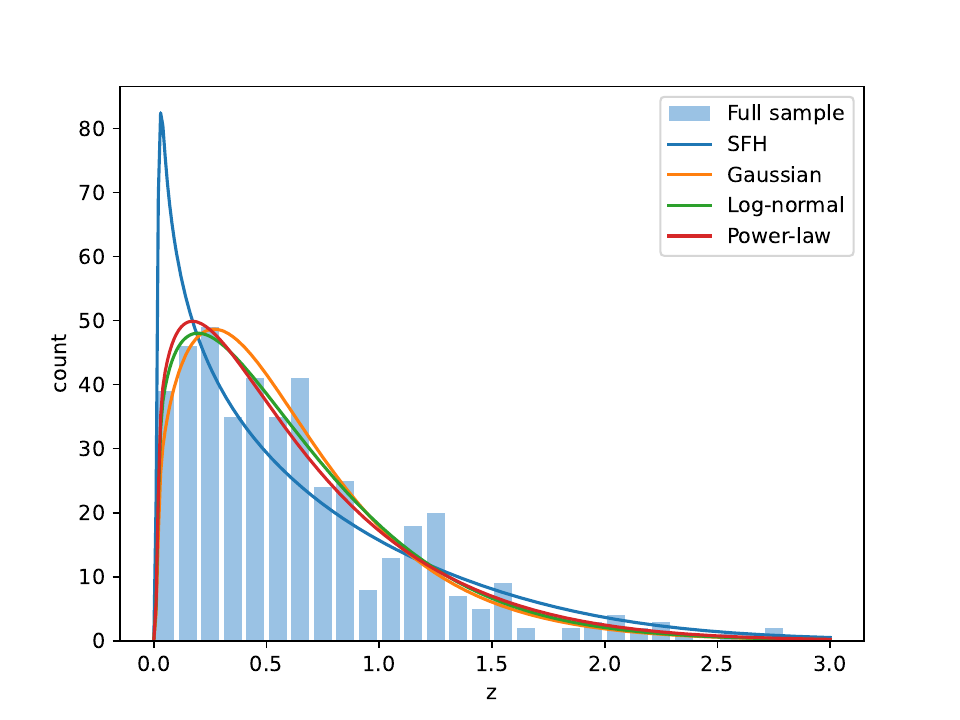}
	\includegraphics[width=0.32\textwidth]{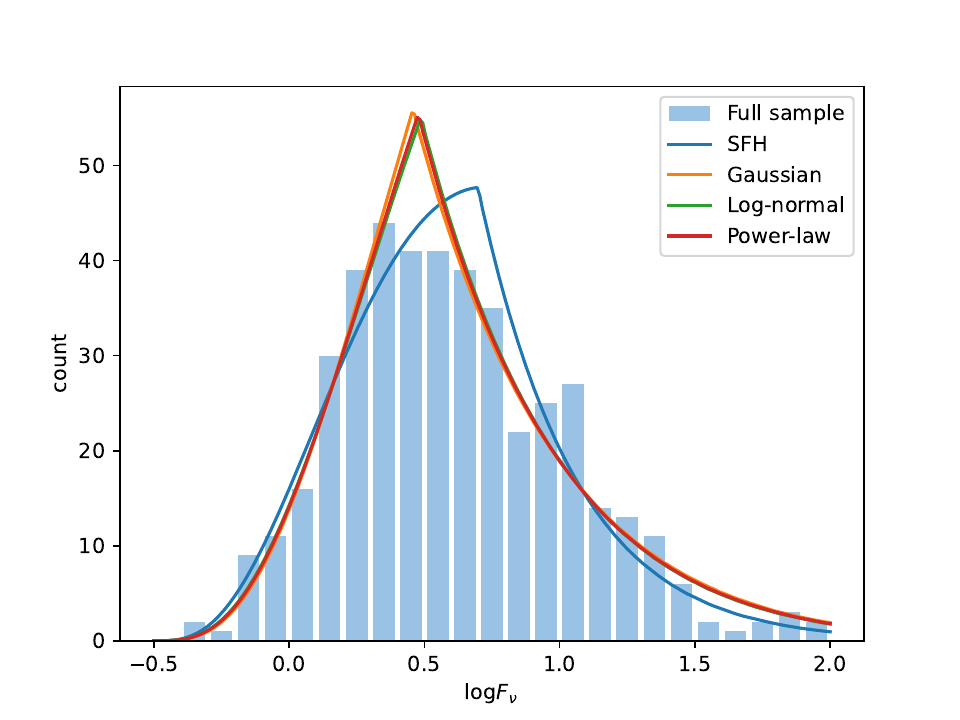}	
    \caption{The histograms of isotropic energy ($\log E$), inferred redshift ($z$) and fluence ($\log F_\nu$) of the Gold sample (upper three panels) and the Full sample (lower three panels). The colored lines illustrate the best-fitting results of the four models considered in this paper, with the corresponding model parameters listed in Table \ref{tab_Gold} and Table \ref{tab_Full}.}\label{fig:hist_chime}
\end{figure}

In addition, we employ the criteria proposed in \citet{CHIMEFRB:2021srp} to filter the data that may be unsuitable for population analyses: (1) signal to noise ratio $S/N > 12$; (2) $\rm DM_{obs} > 1.5max(DM_{NE2001}, DM_{YMW16})$; (3) not detected in far sidelobes; (4) not detected during non-nominal telescope operations; (5) scattering time $\tau_{\rm scat}< $10 ms. The 236 FRBs satisfying these criteria are called the ``Gold sample". For comparison, the Full sample and the Gold sample are used to constrain the free parameters separately. We uniformly divide redshift $z$ into 30 bins in the range $(0,3)$, divide fluence $\log F_\nu$ (in unit of Jy ms) into 25 bins in the range ($-0.5,2$), and divide energy $\log E$ (in unit of erg) into 30 bins in the range $(37,43)$. The histograms of energy, redshift and fluence for the Gold sample and the Full sample are shown in Figure \ref{fig:hist_chime}. As is seen, there is no significant difference between this two samples.

\begin{figure}[htbp]
	\centering
	\includegraphics[width=0.48\textwidth]{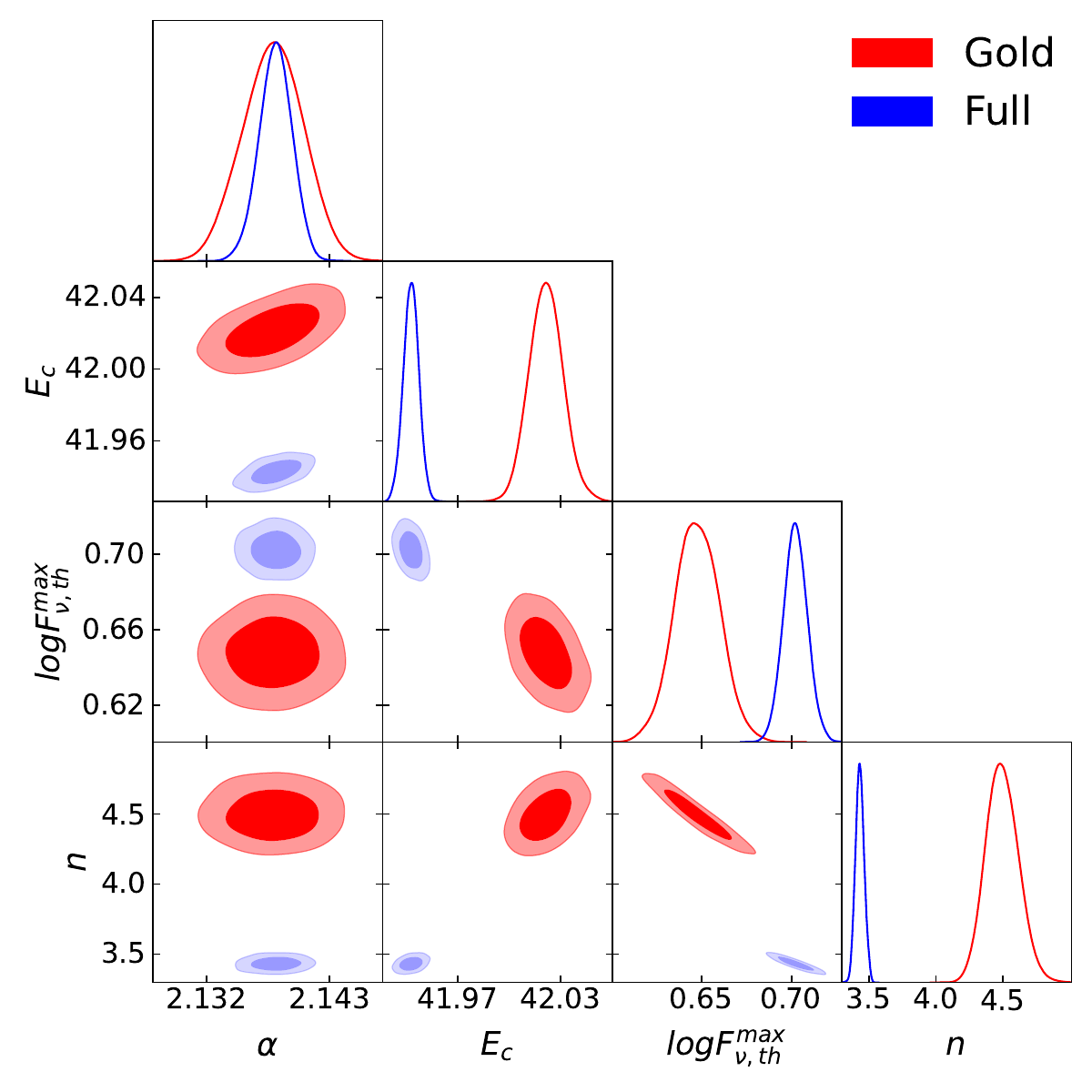}
	\includegraphics[width=0.48\textwidth]{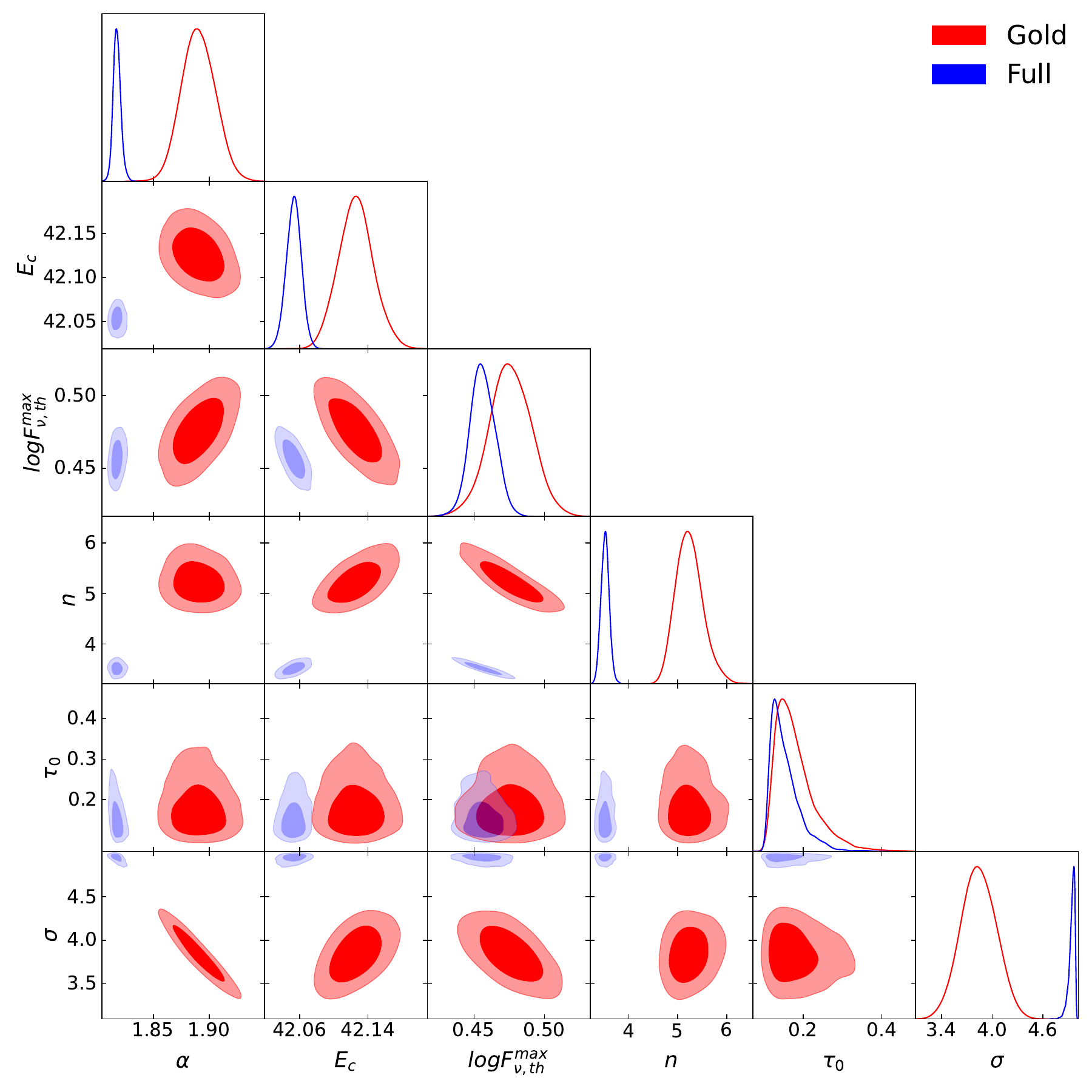}
	\includegraphics[width=0.48\textwidth]{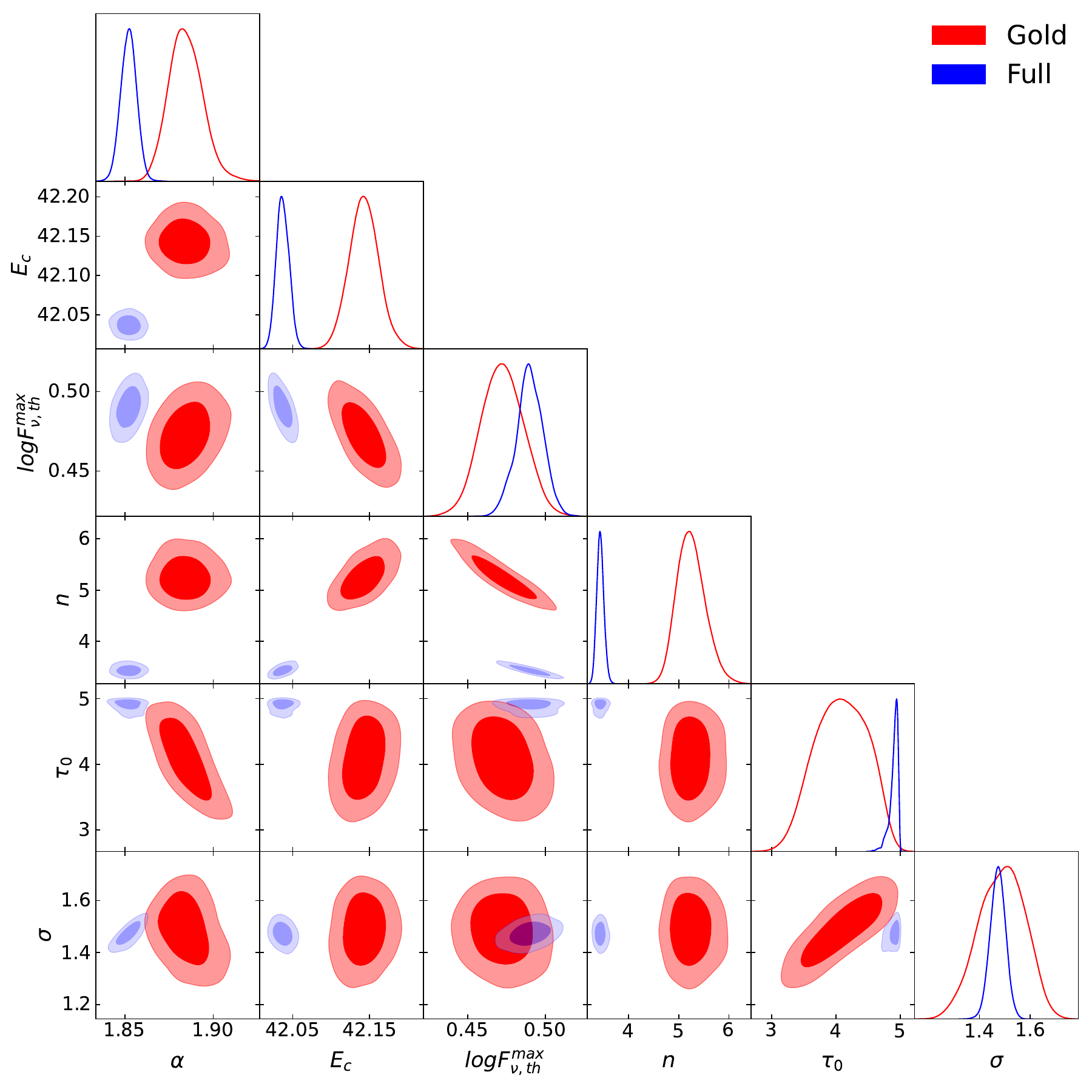}
    \includegraphics[width=0.48\textwidth]{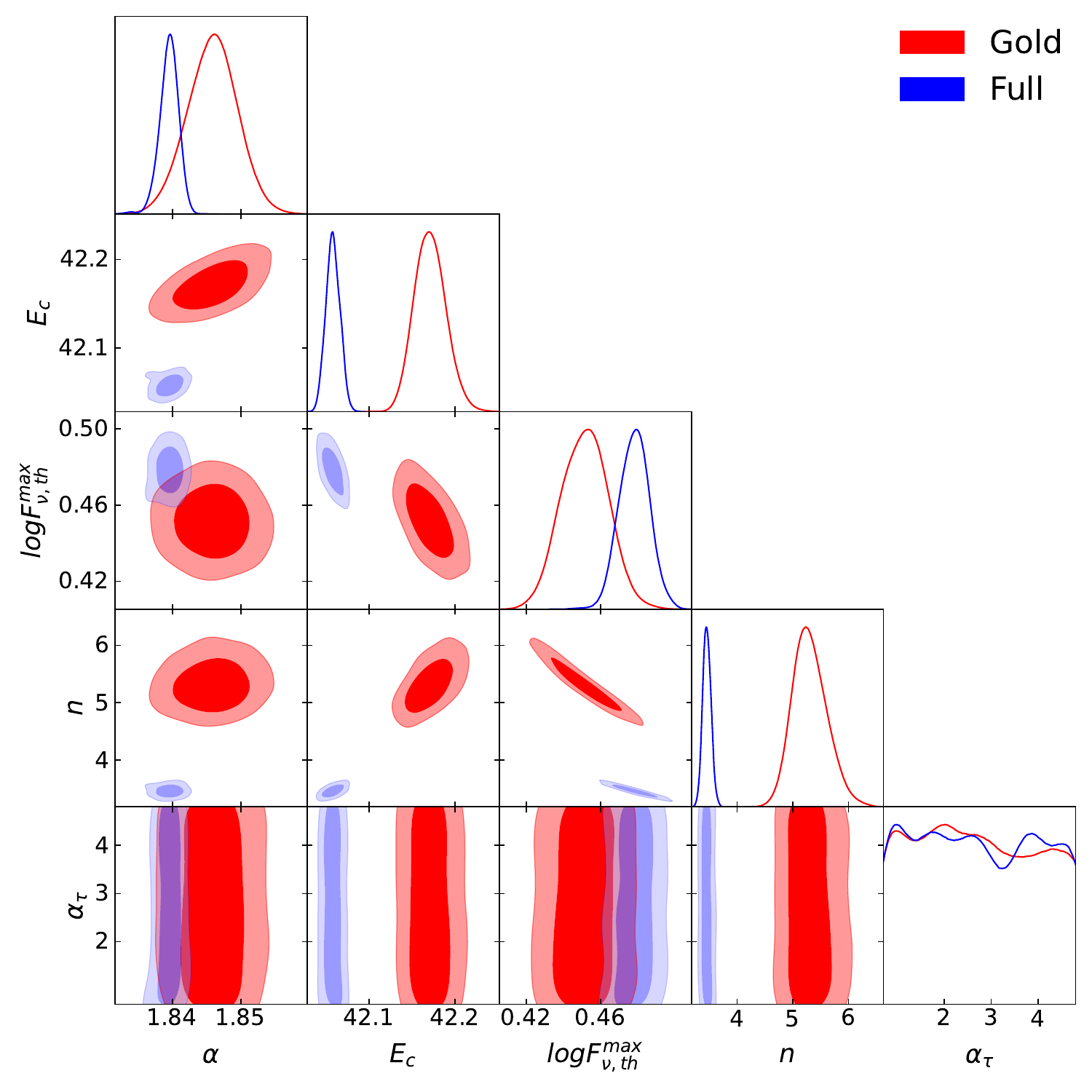}
    \caption{The contour plots and marginalized posterior PDFs of the parameter space. Upper-left: SFH model; upper-right: Gaussian delay model; lower-left: log-normal delay model; lower-right: power-law delay model.}\label{fig:contour}
\end{figure}

Based on the likelihood given in equation (\ref{eq:likelihood}), we calculate the posterior PDFs of the free parameters using the Python package emcee \citep{ForemanMackey:2012ig}. We take non-informative priors for all the free parameters in a wide range, specifically, $\alpha \in U(0.1,5)$, $\mathrm{log}E_c \in U(37,44)$, $n \in U(0.1,8)$, $\mathrm{log}F_{\nu,\rm th}^{\rm max} \in U(0.01,2.0)$, $\tau_0\in U(0.1,5)$, $\sigma \in U(0.1,5)$, $\alpha_\tau \in U(0.1,5)$, where $\tau_0$ and $\sigma$ are in unit of billion years (Gyr). The optimal parameters (the median values of the posterior PDFs) and their 1$\sigma$ uncertainties are summarized in Table \ref{tab_Gold} for the Gold sample, and in Table \ref{tab_Full} for the Full sample. The best-fitting curves are over-plotted in the histograms in Figure \ref{fig:hist_chime}. The posterior PDFs of the model parameters and the two-dimensional confidence contours are shown in Figure \ref{fig:contour}. Note that the best-fitting parameters and best-fitting curves are derived using the CDF rather than the PDF, so they are independent of the binning of data points. We plot the best-fitting curves in the histogram just for visualization purpose, because curves of different models have significant overlap in the CDF plot.

\begin{table}[htbp]
    \centering
    \caption{\small{The optimal parameters and their 1$\sigma$ uncertainties constrained from the Gold sample. The parameters $\tau_0$ and $\sigma$ are in unit of billion years (Gyr).}}\label{tab_Gold}
    \renewcommand\arraystretch{1.5}
    \begin{tabular}{c|cc|cc|cc|c}
    \hline
    Model & $\alpha$ & log$E_{c}$/erg & log$F_{\nu,\rm th}^{\rm max}$ & $n$ & Model parameter & \hspace{1em} & $\Delta$BIC \\
    \hline
    SFH & $2.14^{+0.01}_{-0.01}$ & $42.02^{+0.01}_{-0.01}$ & $0.65^{+0.01}_{-0.01}$ & $4.49^{+0.13}_{-0.12}$ & N/A & N/A & 588.64 \\
    Gaussian & $1.89^{+0.02}_{-0.01}$ & $42.13^{+0.02}_{-0.02}$ & $0.48^{+0.02}_{-0.01}$ & $5.22^{+0.28}_{-0.27}$ & $\tau_0 = 0.16^{+0.09}_{-0.04}$ & $\sigma = 3.84^{+0.22}_{-0.21}$ & 26.61 \\
    log-normal  & $1.88^{+0.01}_{-0.01}$ & $42.14^{+0.02}_{-0.02}$ & $0.47^{+0.01}_{-0.01}$ & $5.23^{+0.30}_{-0.27}$ & $\tau_0 = 4.09^{+0.58}_{-0.55}$  & $\sigma = 1.49^{+0.09}_{-0.10}$  & 0(fiducial)\\
    power-law  & $1.85^{+0.01}_{-0.01}$ & $42.17^{+0.02}_{-0.02}$ & $0.45^{+0.01}_{-0.01}$ & $5.28^{+0.34}_{-0.28}$ & $\alpha_\tau = 2.72^{+1.85}_{-1.68}$& \hspace{1em} & 12.43\\
    \hline
    \end{tabular}
\end{table}

\begin{table}[htbp]
    \centering
    \caption{\small{The optimal parameters and their 1$\sigma$ uncertainties constrained from the Full sample. The parameters $\tau_0$ and $\sigma$ are in unit of billion years (Gyr).}}\label{tab_Full}
	\renewcommand\arraystretch{1.5}
	\begin{tabular}{c|cc|cc|cc|c}
	\hline
	Model & $\alpha$ & log$E_{c}$/erg & log$F_{\nu,\rm th}^{\rm max}$ & $n$ & Model parameter & \hspace{1em} & $\Delta$BIC \\
	\hline
	SFH & $2.14^{+0.01}_{-0.01}$ & $41.94^{+0.01}_{-0.01}$ & $0.70^{+0.01}_{-0.01}$ & $3.43^{+0.03}_{-0.03}$ & N/A & N/A &2015.58 \\
    Gaussian & $1.82^{+0.01}_{-0.01}$ & $42.05^{+0.01}_{-0.01}$ & $0.46^{+0.01}_{-0.01}$ & $3.52^{+0.08}_{-0.08}$ & $\tau_0 = 0.14^{+0.06}_{-0.03}$ & $\sigma = 4.96^{+0.03}_{-0.06}$  & 56.28\\
    log-normal  & $1.85^{+0.01}_{-0.01}$ & $42.04^{+0.01}_{-0.01}$ & $0.49^{+0.01}_{-0.01}$ & $3.43^{+0.07}_{-0.07}$ & $\tau_0 = 4.93^{+0.05}_{-0.11}$  & $\sigma = 1.47^{+0.03}_{-0.03}$  &  0(fiducial)\\
    power-law  & $1.84^{+0.01}_{-0.01}$ & $42.06^{+0.01}_{-0.01}$ & $0.48^{+0.01}_{-0.01}$ & $3.46^{+0.08}_{-0.07}$ & $\alpha_\tau = 2.66^{+1.96}_{-1.81}$& \hspace{1em} & 5.97\\
    \hline
	\end{tabular}
\end{table}

From Table \ref{tab_Gold}, Table \ref{tab_Full} and Figure \ref{fig:contour}, we can see that all parameters, except for the power-law index $\alpha_\tau$ in the power-law delay model, can be tightly constrained. The selection function strongly depends on the FRB samples. Specifically, the parameter $n$ constrained from the Gold sample ($n\sim 5.2$) is much larger than that constrained from the Full sample ($n\sim 3.5$). The parameters related to the energy distribution and redshift distribution are almost independent of the samples, implying that our method is robust. The energy function parameters are consistent in three time-delayed models, with $\alpha\sim 1.8$ and $\log E_c \sim 42$. The SFH model gives a little larger value of $\alpha\sim 2.1$, but the cutoff energy $\log E_c$ is consistent with the three time-delayed models. In the log-normal delay model, the median time delay is $\tau_0=4.09_{-0.55}^{+0.58}$ Gyr for the Gold sample, and $\tau_0=4.93_{-0.11}^{+0.05}$ Gyr. The Gaussian delay model gives a much smaller value of $\tau_0$, i.e. $\tau_0=0.16_{-0.04}^{+0.09}$ Gyr for the Gold sample, and $\tau_0=0.14_{-0.03}^{+0.06}$ Gyr for the Full sample. But be caution that $\tau_0$ is not the average time delay, because the time delay is positive definite. The average time delay for the Gaussian delay model can be calculated by $\bar\tau=\int_0^\infty \tau f(\tau)d\tau/\int_0^\infty f(\tau)d\tau$. Thus we obtain $\bar\tau\sim 3.12$ Gyr for the Gold sample and $\bar\tau\sim 3.95$ Gyr for the Full sample, which are comparable with the median time delays of the log-normal delay model. As for the power-law delay model, the power-law index can't be well constrained, so we can't give a reasonable estimation on the average time delay.

To quantify the goodness of fit and pick up the optimal model, we calculate the Bayesian information criterion (BIC), which is defined by \citep{Schwarz1978}
\begin{equation}\label{eq_Bic}
	{\rm BIC} = -2 {\rm ln}\mathcal{L}_{\rm max}+ k {\rm ln}(N),
\end{equation}
where $\mathcal{L}_{\rm max}$ represents the maximum likelihood estimated at the best-fitting parameters, $k$ denotes the number of free parameters, and $N$ is the total number of data points. A model with a smaller BIC value indicates a superior fit. Note that it is the relative value of BIC, rather than its absolute value, that holds significance. Choosing the model which has the smallest BIC value as the fiducial model, the relative value of BIC is defined as
\begin{equation}\label{eq_deltaBIC}
    \Delta \rm BIC_{model} = BIC_{model} - BIC_{fiducial}.
\end{equation}
According to the Jeffreys' scale \citep{Jeffreys1998,Liddle_2007}, a model with $\rm \Delta BIC \ge 5$ or $\rm \Delta BIC \ge 10$ indicates that there is `strong' or `decisive' evidence for disfavoring this model against the fiducial model.

In the last column of Table \ref{tab_Gold} and Table \ref{tab_Full}, we list the $\rm \Delta BIC$ value of each model. We can see that for
both data samples, the model superiorities are in the same order, namely Log-normal $>$ Power-law $>$ Gaussian $>$ SFH. Both samples most prefer the log-normal model, so it is chosen as the fiducial model. According to the $\Delta{\rm BIC}$ value, the power-law model and the Gaussian model fit the data worse than the log-normal model, but they are much better than the SFH model. Compared with the three time-delayed models, the SFH model is decisively ruled out. This can also be clearly seen from Figure \ref{fig:hist_chime}, in which the SHF curves significantly deviate from the data points. In summary, the SFH model is decisively excluded by both samples, but all the three time-delayed models can to some extent match both samples.

\section{Discussion and Conclusions}\label{sec:conclusions}

In this paper, we investigated the FRB population using the first CHIME/FRB catalog. The ${\rm DM_E}-z$ relation was used to infer redshift and energy of each FRB. We constructed a Bayesian framework to fit the models simultaneously to the distributions of energy, redshift and fluence. This approach allows us to derive the best-fitting parameters when taking the selection effect into account. The observables of FRBs are mainly influenced by three factors: the intrinsic redshift distribution, the intrinsic energy distribution, and the selection effect of detector. The intrinsic energy distribution is modeled by the cutoff power-law. The selection effect is modeled by a two-parametric function of fluence. The intrinsic redshift distribution is what we are most interested in, and four models are considered, namely the SFH model and three time-delayed models. For comparison, two FRB samples are used to fit the models separately.

We constructed the joint likelihood of energy, redshift and fluence, and calculate the posterior PDFs of the parameters using the Bayesian inference method. The best-fitting parameters are summarized in Table \ref{tab_Gold} for the Gold sample and Table \ref{tab_Full} for the Full sample. Different samples mainly affect the selection function, but the model parameters are insensitive to the samples, implying the robustness of our method. The energy function is almost independent of the redshift model we chosen. The best-fitting power-law index $\alpha\approx 1.8$ is in good agreement with the previous results \citep{Lu_2019,Luo:2018tiy,Luo:2020wfx,lu2020unified,lin2023revised}. In addition, the cutoff energy is also tightly constrained to be $\log E \approx 42$, which is consistent with the results of \citet{lin2023revised}, but it is slightly larger than the results of \citet{zhang2022chime} and \citet{Qiang:2021ljr}. Both FRB samples support the log-normal model as the best model, but the other two time-delayed models (Gaussian model and power-law model) can also match the data to some extent. Compared with the three time-delayed models, the SFH model is conclusively ruled out. In summary, there is no doubt that the FRB population has significant time delay with respective to the SFH, but the distribution of time delay remains to be further investigated.

The time-delayed models have already been exhaustively studied using the first CHIME/FRB catalog by \citet{zhang2022chime}. There are two main differences between our work and \citet{zhang2022chime}. The first important difference is the method used to infer redshift. In our work, we reconstruct the ${\rm DM_E}-z$ relation using Bayesian inference method, and the probability distributions of ${\rm DM_{IGM}}$ and ${\rm DM_{host}}$ are considered. On the contrary, \citet{zhang2022chime} assumed that ${\rm DM_{host}}$ is a constant, and directly calculated redshift by solving equation (\ref{eq:DM_IGM}). As is shown in \citet{Tang_2023}, the ${\rm DM_E}-z$ relation curve obtained using Bayesian method is relatively lower than that directly calculated using equation (\ref{eq:DM_IGM}). Hence the inferred redshift (so the energy) we obtained here is larger than that in \citet{zhang2022chime}. This is one of the reasons why the cutoff energy we obtained is larger than that of \citet{zhang2022chime} (another reason is that \citet{zhang2022chime} does not consider the k-correction in energy calculation). The second major difference is the method of parameter inference. In our work, we construct a joint likelihood of redshift, energy and fluence, and optimize the parameters using the Bayesian inference method. On the contrary, \citet{zhang2022chime} divided the parameter space into grids, and use the KS-test to check if each set of parameters can match the observational data. The Bayesian inference method has advantage over the KS-test method, because the latter can't derive the optimal parameters. In spite of these differences, we reach the same conclusion that the FRB population does not trace the SFH. With the same CHIME/FRB sample, \citet{Hashimoto:2022llm} also showed that the SFH model is ruled out with high confidence level.

Similar work has been done by \citet{Shin:2022crt}, but they found that the SFH model couldn't be ruled out, which is in confliction with ours. The sample used by \citet{Shin:2022crt} has significant overlap with the Gold sample used in our paper. So the different conclusions are unlikely to arise from the data samples, but may be caused by the different methods we used. In our paper, we first use the well-localized FRBs to reconstruct the ${\rm DM_E}-z$ relation, and then use it to infer the redshift and energy of the CHIME/FRBs. In contrast, the well-localized FRBs were not used by \citet{Shin:2022crt}. The parameters related to the ${\rm DM_E}-z$ relation ($\mu_{\rm host}$ and $\sigma_{\rm host}$) were constrained simultaneously with the parameters related to the energy and redshift distributions. They also included the spectral index $\alpha$ as a free parameter. So the parameter space in \citet{Shin:2022crt} is larger than that in our paper, leading to relatively larger uncertainties of parameters. In addition, the redshift distribution models we considered here are different from that of \citet{Shin:2022crt}. In \citet{Shin:2022crt}, the authors parameterized the redshift distribution as $p(z)\propto [{\rm SFH}(z)]^n$. The best-fitting value $n=0.96_{-0.67}^{+0.81}$ is well consistent with unity, so they concluded that the CHIME/FRB population is consistent with SFH. But the uncertainty of $n$ is very large, so the time delay is still possible. Note that the power-law form $p(z)\propto [{\rm SFH}(z)]^n$ is too simple, and the peak redshift (i.e. the redshift at which $p(z)$ reaches its maximum) is independent of $n$. So the power-law parametrization of $p(z)$ is hard to reveal the time delay of FRB population with respect to SFH, even if the time delay really exists.

The uncertainty of our method mainly arises from the inference of redshift. We infer redshift of each FRB using the ${\rm DM_E}-z$ relation, which is reconstructed from 17 well-localized FRBs at low redshift ($z<1$). We assume that the parameters of the ${\rm DM_E}-z$ relation is redshift-independent and directly extrapolate them to high redshift. Due to the lack of high-redshift FRBs, the ${\rm DM_E}-z$ relation may has large uncertainty at $z>1$. \citet{James:2021jbo} pointed out that there is a critical point beyond which the one-to-one correspondence between ${\rm DM_E}$ and redshift may be broken. If this is indeed the case, it is impossible to use the ${\rm DM_E}-z$ relation to infer redshift. When calculate the burst energy, we include a k-correction term to transform it to the rest frame, see equation (\ref{eq:E}). However, the k-correction depends on the power-law index of energy spectra, which has a large uncertainty for the CHIME/FRB catalog. To check if the k-correction will significantly affect our results or not, we recalculate the burst energy without the k-correction, i.e. $E=4\pi d_L^2F_{\nu} \Delta\nu/(1+z)$, and redo our Bayesian analysis. We find that including the k-correction or not can slightly affect the best-fitting parameters, but does not change our main conclusions.

In summary, it is still premature to make a conclusive statement about the FRB population based on the currently available data. We can confidently veto the hypothesis that FRBs trace the SFH. There is strong evidence that the FRB population has on average $3\sim 5$ billion years time delay with respect to the SFH, but the accurate probability distribution of time delay is still unconstrained. The log-normal delay model seems to match the data best, but the power-law delay model and Gaussian delay model are also acceptable. We anticipate that a large sample of well localized FRBs in the near future can provide definitive insights into the FRB population.

\begin{acknowledgements}
This work has been supported by the National Natural Science Fund of China under grant nos. 12275034 and 12347101, and the Fundamental Research Funds for the Central Universities of China under grant no. 2023CDJXY-048.
\end{acknowledgements}

\bibliography{reference}{}
\bibliographystyle{aasjournal}

\end{document}